\documentclass[12pt,preprint]{aastex}

\def\gta{\ifmmode {\mathbin{\lower 3pt\hbox   %> or of order
    {$\,\rlap{\raise 5pt\hbox{$\char'076$}}\mathchar"7218\,$}}}
    \else {${\mathbin{\lower 3pt\hbox
    {$\rlap{\raise 5pt\hbox{$\char'076$}}\mathchar"7218\,$}}}
    $}\fi}
\def\lta{\ifmmode {\,\mathbin{\lower 3pt\hbox   %< or of order
    {$\,\rlap{\raise 5pt\hbox{$\char'074$}}\mathchar"7218\,$}}}
    \else {${\mathbin{\lower 3pt\hbox
    {$\rlap{\raise 5pt\hbox{$\char'074$}}\mathchar"7218\,$}}}
    $}\fi}

\shorttitle {Double-peaked burst from 4U 1636--536}
\shortauthors {Bhattacharyya and Strohmayer}

\begin{document}

\title {A non-PRE double-peaked burst from 4U 1636--536: evidence for
burning front propagation}

\author {Sudip Bhattacharyya\altaffilmark{1,2}, and Tod
E. Strohmayer\altaffilmark{2}}

\altaffiltext{1}{Department of Astronomy, University of Maryland at
College Park, College Park, MD 20742-2421}

\altaffiltext{2}{X-ray Astrophysics Lab,
Exploration of the Universe Division,
NASA's Goddard Space Flight Center,
Greenbelt, MD 20771; sudip@milkyway.gsfc.nasa.gov,
stroh@clarence.gsfc.nasa.gov}

\begin{abstract}
We analyse Rossi X-ray Timing Explorer (RXTE) Proportional Counter
Array (PCA) data of a double-peaked burst from the low mass X-ray
binary (LMXB) 4U 1636--536 that shows no evidence for photospheric
radius expansion (PRE). We find that the X-ray emitting area on the
star increases with time as the burst progresses, even though the
photosphere does not expand. We argue that this is a strong indication
of thermonuclear flame spreading on the stellar surface during such
bursts. We propose a model for such double-peaked bursts, based on
thermonuclear flame spreading, that can qualitatively explain their
essential features, as well as the rarity of these bursts.
\end{abstract}

\keywords{accretion, accretion disks --- relativity --- stars: neutron --- 
X-rays: binaries --- X-rays: bursts ---  X-rays: individual (4U 1636--536)}

\section {Introduction} \label{sec: 1}

X-ray bursts are produced by thermonuclear burning of matter
accumulated on the surfaces of accreting neutron stars (Grindlay et
al. 1976; Belian, Conner, \& Evans 1976; Woosley \& Taam 1976, Joss
1977; Lamb \& Lamb 1978). For most bursts, profiles are single
peaked, with rise times of the order of a fraction of a second to a
few seconds, and decay times of the order of ten or a few tens of
seconds.  However, for some bursts, double-peaked structure is
observed.  These peaks (with time separation of a few seconds) in a
single luminous burst can normally be explained in terms of
photospheric radius expansion (PRE; due to radiation pressure) and
contraction (Paczynski 1983; Ebisuzaki, Hanawa, \& Sugimoto 1984). As
the photosphere expands, the effective temperature decreases, and the
emitted photons shift towards lower energies. A subsequent contraction
of the photosphere has the opposite effect. This can cause a dip (and
hence the double-peaked structure) in the high-energy burst profile
(Lewin et al. 1976; Hoffman, Cominsky, \& Lewin 1980), although such a
structure is not frequently seen in bolometric or low-energy profiles
(see Smale 2001).

Double-peaked structure in weak X-ray bursts was discovered by Sztajno
et al. (1985) using EXOSAT observations of the low mass X-ray binary
(LMXB) system 4U 1636--536. For these bursts, two peaks are seen in
the bolometric profile, and even in low-energy profiles.  For this
reason, and as these bursts are not strong enough to cause
photospheric expansion, some other physical effects are needed to
explain them. Several models have been put forward to explain these
non-PRE double-peaked bursts: (1) two-step energy generation due to
shear instabilities in the fuel on the stellar surface (Fujimoto et
al. 1988), (2) a nuclear waiting point impedance in the thermonuclear
reaction flow (Fisker, Thielemann, \& Wiescher 2004), (3) heat
transport impedance in a two-zone model (Regev \& Livio 1984), and
(4) interactions with the accretion disk (Melia \& Zylstra 1992). As
we will elaborate in \S~3, none of these models can explain various
aspects of these bursts satisfactorily.
In this Letter, we propose a model for the double-peaked bursts based on thermonuclear flame
spreading on neutron stars, and 
comparing it qualitatively with the RXTE data of a double-peaked burst from
4U 1636--536, we show that our model can explain the essential features of these bursts.

\section {Data Analysis and Model Calculations} \label{sec: 2}

We analyse the RXTE PCA archival data of a double-peaked burst (Date
of observation: Jan 8, 2002; ObsId: 60032-01-19-000) from 4U
1636--536.  The heights of the two peaks are almost identical $(\sim
2200$ counts/s/PCU), with a dip depth more than half
the peak height (Fig. 1).  This is a weak burst
compared to PRE bursts from this source, which can have $\sim 7000$
counts/s/PCU (see Strohmayer et al. 1998).  The burst
profiles at different energy bands are very similar
(Fig. 1), showing that this is not a PRE burst. However, the hardness
in panel {\it b} of Fig. 1 shows two striking features: (1) the first
peak of the hardness occurs $2-3$ seconds before that of the burst
profile; (2) the second hardness peak is much lower than the first
one, while the burst profile peaks are of similar height. As the
emitted flux primarily depends on source hardness (which is a measure
of temperature) and source emission area, feature (1) indicates that
the emission area increases with time. Feature (2) is possible if the
emission area at the time of the second peak is much higher than that
at the time of the first peak. As for a non-PRE burst, the emission area
can increase only if the burning region spreads on the stellar surface
from an initially small size, these two features
are consistent with thermonuclear flame spreading (Strohmayer, Zhang, \& Swank 
1997; Kong et al. 2000).

As a next step, we break the burst profile into smaller time bins, and
for each bin perform spectral fitting. This gives the time evolution
of the spectral parameters. We fit the data with a single temperature
blackbody model (bbodyrad in XSPEC), as generally burst spectra are
well fit by a blackbody (Strohmayer \& Bildsten 2003). In doing this,
we fix the hydrogen column density $N_{\rm H}$ at a value
$0.56\times10^{22}$~cm$^{-2}$ (van Paradijs et al. 1986).
The results of these fits are shown in panels {\it c} \& {\it d} of
Fig. 1.  The radius is calculated from the ``normalization'' and
provides a measure of the source emitting area. The panels show that
the evolution of the temperature is similar to that of the hardness
(as expected), and the size of the emission area increases with time
(indicating flame spreading), first quickly, and then more slowly. The
temporal behavior of the radius also shows that this is not a PRE
burst, otherwise the radius would decrease from the time when the
burst profile attains its minimum between the two peaks.  However, the
reduced $\chi^2$ values are high for these fits $(> 1.5$ 
for 13 out of 29 time bins).  Considering
the arguments of the previous paragraph, this may be because of the
following reason: the emission is locally blackbody, but temperatures
at different locations on the stellar surface are significantly
different (as a result of slow flame spreading in comparison to the
timescale of temperature decay at a given location), and hence a
single temperature blackbody model can not fit the observed spectra
well.  However, the similar evolution of temperature to that of the
hardness indicates that these fits give
average blackbody temperatures on the stellar surface. This
explains the smaller height of the second temperature
peak (panel {\it c}, Fig. 1), as with the slow flame
spreading, temperature decays on most part of the star
before the flame engulfs the whole star, making the average
temperature smaller during the second peak.
The error bars in panels {\it c} \& {\it d} of Fig. 1 give $1\sigma$
errors.  As the reduced $\chi^2$ values for some of the time bins are high, 
increasing $\chi^2$ by 1 from the best fit value would
underestimate these errors. Therefore, we increase $\chi^2$ by
the amount of the reduced $\chi^2$ of the fit to calculate the
$1\sigma$ errors.

From the above analysis we infer that double-peaked bursts may be
caused by thermonuclear flame spreading on the stellar surface.
In order to show this, in our simple model, we consider that the fuel
(accreted matter) is distributed over the entire stellar surface,
the burst is ignited at a certain point, 
and then propagates on the surface to ignite all the fuel
gradually. For the particular double-peaked burst
analysed here, we assume that 
the burning region forms a $\phi$-symmetric belt very
quickly after ignition at or near the north pole
(while the observer's inclination angle, measured from this pole,
is $\le 90^{\rm o})$, in order to explain the non-observation
of millisecond period brightness oscillations, and the initially fast moving
front ``stalls'' for a time as it approaches the equator, before
speeding up again into the opposite hemisphere. 
This causes the burning front to take more time to reach the equator from
the mid-latitudes, and during that time hot portions of the star can
cool, causing a decrease in the emitted flux. Approaching the
equator, the front propagation speed increases again,
causing an increase of the emitted flux and the
observed double-peaked structure.

To qualitatively test this hypothesis we calculate the corresponding
model, assuming that the emitting region is a $\phi$-symmetric
belt extending from the north pole to a polar angle $\theta_{\rm edge}$.
To compare this model with the data, we need
to calculate the flux and spectrum at a certain time elapsed since burst onset
$(\Delta t)$, and hence it is essential to know $\theta_{\rm edge}$ and the temperature
at a given $\theta$-position in the belt as functions of $\Delta t$.
The first one can be determined from 
$\Delta t = \int_0^{\theta_{\rm edge}} d\theta/\dot\theta(\theta)$, 
if the burning front speed $\dot\theta(\theta)$ is known. To calculate
the temperature, we assume that after ignition the temperature
increases from $T_{\rm low}$ to 
$(T_{\rm low} + (0.99\times (T_{\rm high} - T_{\rm low})))$
following the equation $T(t) = T_{\rm low} + (T_{\rm high} - T_{\rm
low})\times(1-\exp(-t/t_{\rm rise}))$, and then decays exponentially with an
e-folding time $t_{\rm decay}$. In our model, we assume that
$\dot\theta(\theta) = F(\theta) = 1/(t_{\rm total}\times\cos\theta)$
for $\theta \le 90^{\rm o}$, and $\dot\theta(\theta) = F(180^{\rm o}-\theta)$
for $\theta \ge 90^{\rm o}$, where 
$t_{\rm total}$ is the time needed by the front to propagate from a pole to
the equator in the absence of any stalling. 
This expression of $\dot\theta(\theta)$ 
follows from Spitkovsky et al. (2002), as the neutron star in 4U 1636--536 is
rapidly rotating (spin frequency $\nu_* \approx 582$~Hz; Giles et al. 2002;
Strohmayer \& Markwardt 2002), and hence the effect of
the Coriolis force on the flame speed should be important. We assume that the
stalling of the front happens between the polar angles $\theta_1$ and $\theta_2$
in the northern hemisphere: $\dot\theta(\theta)$ decreases linearly from $\theta = \theta_1$ to
$\theta = \theta_m$, reaching a value $s/t_{\rm total}$, and then
increases linearly up to $\theta = \theta_2$ reaching a value $F(\theta_2)$.
In our calculations, we consider the Doppler, special relativistic, and
general relativistic (gravitational redshift and light-bending in
Schwarzschild spacetime) effects.
We compute model lightcurves and spectra for a range of parameter
values, and show an example in Fig. 2, which qualitatively reproduces
the observed features of the double-peaked burst. 
In panel {\it a} of Fig. 2, the burst profiles
qualitatively match (including the
depth of the dip) the data (see Fig. 1), except the
initial rise. For the model, the initial rise time is longer than that
for the data. An effect which may account for this discrepancy is the
radiative diffusion time, ie. the delay between ignition at depth and
emergence of the radiation.  Note also that we calculate the model
flux only up to the time when the burning front reaches the south
pole, while in Fig. 1, the real data probably extend beyond that time.
In panels {\it b} and {\it c} of
Fig. 2, we plot the model hardness and average temperature on the
stellar surface, respectively. We also fit our normalised model
spectra with the XSPEC model bbodyrad, in the same manner as for the
data. The resulting blackbody temperature and radius are shown in
panels {\it d} and {\it e} of Fig. 2.  Panels {\it b}, {\it c} and
{\it d} show a similar temporal behavior: both hardness and
temperature increase at the beginning rapidly, then decrease up to the
point when the burst profile reaches a minimum, increase slightly up
to the point when the burst profile reaches the second peak, and then
decrease again. This behavior is strikingly similar to that seen in
the burst data (Fig. 1). We note that
the temporal behavior of the model average temperature 
(Fig. 2) suggests that spectral fitting with a single temperature
blackbody model actually does give the average temperature on the
stellar surface. In panel {\it e} of Fig. 2, the evolution of the
radius shows an initial rapid increase, and then a slower increase,
which is also quite similar to the data (Fig. 1).
Therefore, simple modeling of pole to pole flame spreading (with a temporary stalling)
can reproduce the essential features of this double-peaked burst.

We note that the burst lightcurves are sensitive to the values of the parameters,
such as $\theta_1$, $\theta_2$, $\theta_{\rm m}$, etc. quantitatively, 
but not qualitatively. For example, the main effects of the increase of 
$\theta_1$, $\theta_2$, $\theta_{\rm m}$, $s$, $t_{\rm total}$, $t_{\rm rise}$ 
and $t_{\rm decay}$ are 
to decrease $d$, slightly decrease $d$, decrease $p$, decrease $l$,
increase the timescale of the whole burst, increase the rise time of the second peak,
and decrease $l$ respectively. Here,
$d$ is the time separation between the two peaks relative to the burst duration, 
$p$ is the ratio of the flux of second peak to that of first peak, and 
$l$ is the ratio of the depth of the dip (from the second peak) to the second peak
height.

\section {Discussion and Conclusions}

In this Letter we have presented a new model for double-peaked bursts,
that naturally explains the observed increase in emission area, 
which other models do not.  Moreover, it appears unlikely that model 1 (see
\S~1) can reproduce both the burst profile and the evolution
of hardness (or, temperature) simultaneously, as it does not consider
the emission area increase. There is also no real calculation of
double-peaked profiles from this model. In addition, if thermonuclear
flames spread in the way Spitkovsky et al. (2002) argue, it is very
difficult to see how sufficient unburnt fuel (as required by
model 1) can be maintained on top of the burnt fuel, as the full scale
height of the hot fuel is likely overturned and mixed with the cold
fuel.  We suggest that models 2 \& 3 (see \S~1) are probably unable to
reproduce the large dip (judging from the figures of Regev \& Livio
1984; Fisker, Thielemann, \& Wiescher 2004), seen in the observed
burst. It is also unclear whether these models, as well as model 4
(see \S~1), can explain the observed hardness and/or temperature
evolution, and the rarity of the double-peaked
bursts. Note that the naive interpretation of a double-peaked burst
as two subsequent bursts (possibly in two hemispheres) can be ruled
out, because these two bursts would have to be localized (probably
by two magnetic poles; otherwise flame spreading without stalling
would give rise to a single peak), and in that case we would expect to observe
millisecond period brightness oscillations.
Our model can qualitatively reproduce the essential features of the
double-peaked bursts (see \S~2), including the burst profile (with a
large dip) and the hardness evolution. 
However it requires the burning front to stall for a few seconds, which
clearly warrants some justification and further study.
We suggest that accretion may provide a
mechanism to slow the front, although a detailed theoretical
calculation and modeling of the data are required to establish this.
The magnetic field of the neutron star in 4U 1636-53 is
probably comparatively low (as the source is not a millisecond X-ray pulsar), 
and accretion likely proceeds via a disk
around the equatorial plane. 
Therefore, a weak burst ignited near the north pole and proceeding towards
the equator may be impeded and stalled by the pole-ward flow of accreted matter in
the mid-latitudes, as this matter spreads from the equator towards
the poles, first rapidly, then more slowly (Inogamov \& Sunyaev 1999).
After reaching the vicinity of the equator, the burning region may be able to
inhibit accretion sufficiently to allow the front to speed up
again. This is because  the gravitational force on
particles falling onto the star via a disk is closely balanced by the centrifugal
force, and hence even weak bursts can probably inhibit accretion (see Inogamov \& Sunyaev
1999), if thermonuclear flux is radiated near the equator. The cessation of accretion
may also allow the observer to get X-ray flux from the burning region
in some parts of the southern hemisphere. We emphasize that the credibility of these
arguments depends on the justification of accretion induced impedance of
front propagation, as the accretion flows exist much above the burning layer.
Here we give the following qualitative arguments. The $\phi$ component of linear speed 
$(v_{\phi})$ of accreted matter in the stellar atmosphere (in mid-latitudes and near
the equator) is $\sim 10^{5}$ km s$^{-1}$
(Fujimoto et al. 1988), and the corresponding latitudinal $(\theta)$ component 
$(v_{\theta})$ may be $\sim 1000$ km s$^{-1}$ (Inogamov \& Sunyaev 1999). Such 
accreted matter is likely to produce differential rotation in the inner layers
by the inflow of the angular momentum, which may extend down through the
burning shell (at the column depth of $\sim 10^8$ gm cm$^{-2}$; Fujimoto et al. 1988).
Therefore, in the burning layer, $v_{\theta}$ may be comparable to the
flame speed, which may be $\le 10$ km s$^{-1}$ for a rapidly spinning star
$(g \sim 10^{14}$ cm s$^{-2}$, $f = 3657$ rad s$^{-1}$ at $\theta = 60^{\rm o}$;
Spitkovsky et al. 2002). As a result, the burning front could plausibly be influenced
by the accretion-induced pole-ward motion of burning shell matter.

Double-peaked structure appears
only to be associated with weak bursts, perhaps because strong bursts
would tend to disrupt accretion sufficiently to preclude the kind of
front stalling that is required for the occurence of two peaks
according to our model.  The double-peaked feature is somewhat rare
even among the weak bursts.  This may be because in order to have the
double-peaked structure, the burst needs to be ignited at or near a
pole (so that the accretion can continue for a few seconds), which is
less probable than equatorial ignition (Spitkovsky et al. 2002). The
fact that double-peaked bursts are seen from only a few sources
(mostly from 4U 1636--536) can be understood in our model as
follows. These bursts require a low stellar magnetic field (for a
given accretion rate), so that accretion happens mostly through a disk
in the equatorial plane, and the disk must closely approach the star
(so that the gravitational force is closely balanced by the
centrifugal force near the surface). This is possible, if the
stellar equatorial dimensionless radius to mass ratio 
$R/M$ is large, and $\nu_*$ is high (making the radius of
the innermost stable circular orbit small; Bhattacharyya
et al. 2000). This relatively fine tuning among magnetic field,
accretion rate, equatorial $R/M$ and $\nu_*$ may exist for a
relatively small fraction of LMXBs. Therefore, our model
qualitatively explains the enigmatic rarity of the non-PRE
double-peaked bursts, and may also, in principle, enable constraints
on stellar magnetic fields and equatorial $R/M$ to be obtained.

Our work suggests that non-PRE double-peaked bursts can be important
in understanding thermonuclear flame spreading on neutron stars, which
may provide important insights about the millisecond period
brightness oscillations during X-ray bursts, and hence can be useful 
for constraining equation of state models of the dense matter in
the cores of neutron stars (Bhattacharyya \& Strohmayer 2005; 
Bhattacharyya et al. 2005).  However, the
rarity of such bursts has been an obstacle to understanding them, and
thus new attempts to expand the sample of these bursts seems well
warranted.

\acknowledgments

This work was supported in part by NASA Guest Investigator grants.

{}

\clearpage

\begin{figure}
\epsscale{.80}
\hspace{-6.0cm}
\plotone{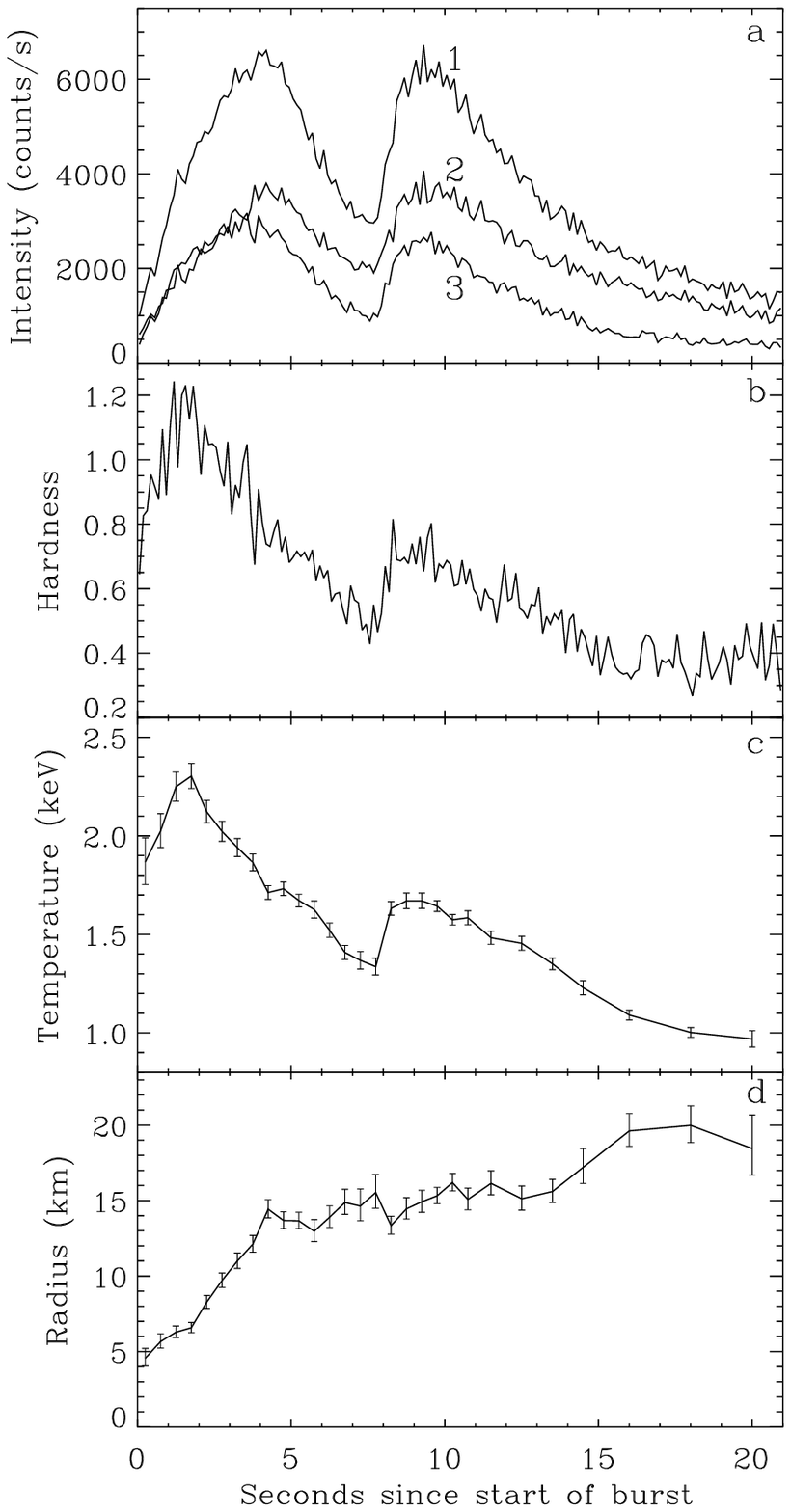}
\vspace{-4.0cm}
\caption{Double-peaked burst from 4U 1636--536: panel {\it a} gives
the burst profiles (for 3 PCUs on): curve 1 is for the channel range $0-63$ (nearly
bolometric), curve 2 is for the channel range $0-10$ (energy $<
6.52$~keV), and curve 3 is for channel range $11-63$ (energy $>
6.52$~keV). Panel {\it b} shows the time evolution of hardness (ratio
of counts in $11-63$ channel range to that in $0-10$ channel
range). For both these panels, the size of the time bin is 0.125 s.
Panels {\it c} \& {\it d} show the time evolution of the blackbody
temperature and the apparent radius (assuming 10 kpc source distance)
of the emission area respectively, obtained by fitting the burst
spectrum (persistent emission subtracted) with a single temperature
blackbody model.} 
\end{figure}

\clearpage

\begin{figure}
\epsscale{.90}
\hspace{-8.0cm}
\plotone{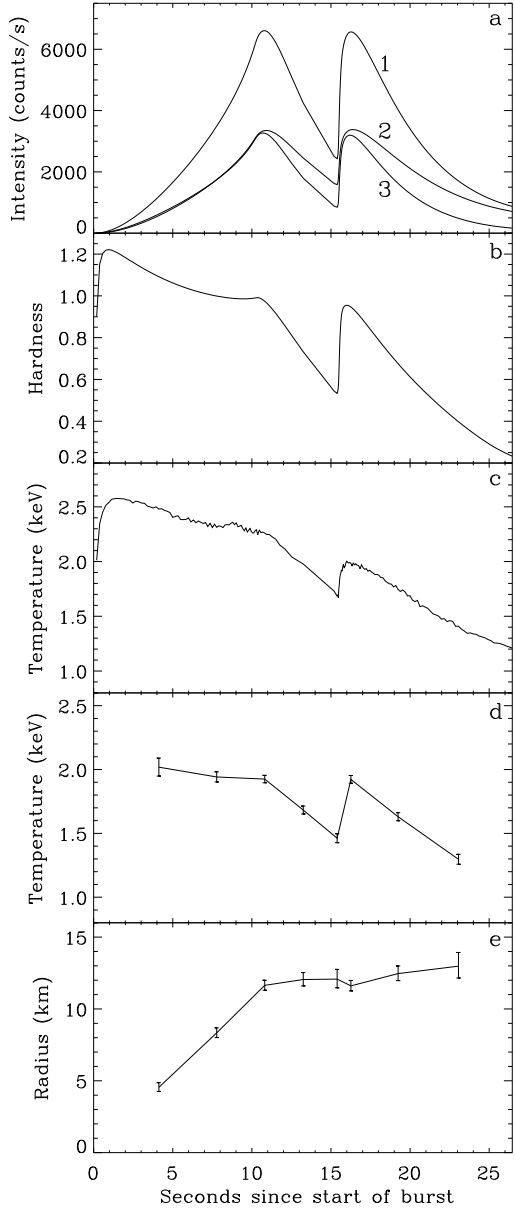}
\vspace{-5.0cm}
\caption{Model (convolved with a PCA response matrix) of
double-peaked bursts: for all the panels, the burst is normalised so
that its first intensity peak has the same count rate as that of the
first peak of the observed burst. Panels {\it a} \& {\it b} are
similar to those of Fig. 1. Panel {\it c} gives the time evolution of
average blackbody temperature on the stellar surface.  Panels {\it d}
\& {\it e} are similar to panels {\it c} \& {\it d} of Fig. 1
respectively. For these two panels, spectra are calculated for 0.5 s
time bins for each point. Model parameter values are the following:
stellar mass $M = 1.5 M_{\odot}$, dimensionless stellar radius to mass
ratio $R/M = 5.5$, stellar spin frequency $\nu_* = 582$~Hz, observer's
inclination angle (measured from north pole) $i = 50^{\rm o}$,
$\theta_1 = 67^{\rm o}$, $\theta_m = 83^{\rm o}$, $\theta_2 = 87^{\rm
o}$, $s = 0.04$, $t_{\rm total} = 11$~s, $t_{\rm rise} = 0.05$~s,
$t_{\rm decay} = 6$~s, $T_{\rm low} = 1$~keV, and $T_{\rm high} =
2.8$~keV (see text for the definitions of the parameters).}
\end{figure}

\end{document}